# Self-training superconducting neuromorphic circuits using reinforcement learning rules

*M. L. Schneider[1], E. M. Jué[2,3], M. R. Pufall[1], K. Segall[4], C. W. Anderson[5]*

1. Applied Physics Division, National Institute of Standards and Technology, Boulder, Colorado 80305

2. Department of Physics, University of Colorado Boulder, Boulder, Colorado 80305

3. Associate of the National Institute of Standards and Technology, Boulder, Colorado 80305

4. Department of Physics and Astronomy, Colgate University, Hamilton, New York 13346

5. Department of Computer Science, Colorado State University, Fort Collins, Colorado, 80523

## Abstract

Reinforcement learning algorithms are used in a wide range of applications, from gaming and robotics to autonomous vehicles. In this paper we describe a set of reinforcement learning-based local weight update rules and their implementation in superconducting hardware. Using SPICE circuit simulations, we implement a small-scale neural network with a learning time of order one nanosecond. This network can be trained to learn new functions simply by changing the target output for a given set of inputs, without the need for any external adjustments to the network. In this implementation the weights are adjusted based on the current state of the overall network response and locally stored information about the previous action. This removes the need to program explicit weight values in these networks, which is one of the primary challenges that analog hardware implementations of neural networks face. The adjustment of weights is based on a global reinforcement signal that obviates the need for circuitry to back-propagate errors.

## Introduction

Neuromorphic computing typically refers to computing hardware whose operation draws inspiration from the way the human brain processes information. [1–4] There are many reasons to look to the brain for computational inspiration. Using roughly 20 Watts of power, the human brain is an extremely energy efficient computational system. In addition, with its highly parallel architecture, the brain is both fault tolerant and can perform certain tasks with surprising speed given the typical neuron firing rate of a few hundred hertz. Perhaps the most intriguing properties of the brain for neuromorphic hardware are its adaptability and capability for learning. Thus, developing self-training hardware stands out as an important goal for neuromorphic systems.

Superconducting hardware lends itself to a neuromorphic approach in part because of the natural spiking behavior of the Josephson junction (JJ), similar to a neuron, and the near lossless propagation of these spikes on superconducting transmission lines, similar to axons. [5] In this paper, we discuss the superconducting implementation and training of a third major element, the synaptic function, which controls the strength of the connection between two neurons. With these basic building blocks, the question becomes how to create a biologically inspired self-training architecture using the elements available that might possibly scale to circuit sizes of technological relevance.

In addition to being naturally neuromorphic, superconducting circuits can operate at speeds in excess of 100 GHz and can be extremely energy efficient. [6–8] For example, the typical spiking energies of the JJs that we are modeling here are less than one attojoule. Once a cryogenic system is scaled beyond the initial research and development phase, it takes less than 1000 W of wall power to cool 1 W of power dissipated at 4 K. This results in spiking energies that are still more than an order of magnitude lower than the human brain's 10 femtojoules per spike, even when accounting for the energy overhead required to operate at 4 K.

To implement a neuromorphic architecture, one needs to consider both the bio-inspired building blocks and the method with which they are trained to perform a certain task. In its most basic form, reinforcement learning establishes a reward function based on the comparison between the desired and actual outputs of a system. [9,10] This function can be written as a set of *local* learning rules with the potential to be efficiently incorporated into neuromorphic hardware systems. The advantage of local learning rules is that they help enable the colocation of memory and logic, using the language of the more traditional von Neumann computational primitives, because they do not rely on detailed cross system (nonlocal) information. We discuss our basic neuromorphic building blocks and the circuit implementation of the reinforcement learning rules in a mixed analog and digital superconducting hardware.

## Materials and methods

### Learning rules

The learning rules defined below for updating the output layer units are the common stochastic gradient descent equations to minimize the mean squared error between the units' outputs, $y_{L,u}$ and their desired outputs, $y^*_{L,u}$. All experiments here are supervised-learning problems for which the desired outputs are known. The common approach for training neural networks to solve supervised learning problems is to use error backpropagation to back-propagate the gradient of the mean squared error with respect to every weight.

To avoid the extra time and circuits required to perform this backwards flow of information through the multiple hidden layers of a neural network, we instead implement a correlative update algorithm involving a global reinforcement signal simultaneously sent to all hidden layer units. While this requires more updates than error backpropagation, it considerably reduces the complexity of the circuits required for training.

The basic definitions and notation used for the network are:

- a network with $L$ layers, each layer $l$ having $N_l$ units

- weights, $w_{l,u,i}$ for layer $l$, unit $u$, and input $i$
- bias weights, $w_{l,u,0}$ for layer $l$, unit $u$
- inputs, $x_{l,i}$ for layer $l$, and input $i$
- weighted sums, $s_{l,u}$ for layer $l$, unit $u$
- output $y_{l,u}$ for layer $l$, unit $u$, which becomes $x_{l+1,u}$, the inputs to the following layer
- stochastic perturbation $d_{l,u}$ for layer $l$, unit $u$, where $d_{l,u} = \{-\delta, 0, \delta\}$
- global reward $r$
- weight update $w_{l,u,i}$ +, with increment size $\rho$
- an input sample consisting of $N_1$ inputs to the first layer, layer 1, with the corresponding desired output $y^*_{L,u}$, for unit $u$ in the last layer, layer L

The equations of the network are slightly different between the hidden layers and the output layer. The output layer is where the global binary reinforcement signal is determined and the target of the output of the units is known directly. The output layer uses its current inputs $x_{L,i}$, outputs $y_{L,u}$ and desired outputs $y^*_{L,u}$ to update its weights. The hidden layers need the current inputs and outputs, a global reinforcement signal, and a memory of the previous output state in order to determine the weight update direction. They also require a stochastic excitation or inhibition input to ensure that the network effectively samples the available states. We find that best practice for learning with this network is for a sample (inputs and desired output) to remain fixed for two time increments before changing conditions. We define the equations of the network as follows:

**Equations for the output layer units**

$$s_{L,u} = w_{L,u,0} + \sum_{i=1}^{N_{L-1}} x_{L,i} w_{L,u,i}$$

$$y_{L,u} = \begin{cases} 1, & \text{if } s_{L,u} > 0 \\ 0, & \text{otherwise} \end{cases}$$

$w_{L,u,i} \mathrel{+}= \rho(y^*_{L,u} - y_{L,u})(x_{L,i} - 0.5)$, for $i > 0$

$w_{L,u,0} \mathrel{+}= \rho(y^*_{L,u} - y_{L,u})$

$$r = \begin{cases} 1, & \text{if network's outputs are all correct} \\ 0, & \text{otherwise} \end{cases}$$

$\Delta r = r^{now} - r^{previous}$

**Equations for hidden layer units**

$x_{l,i} = y_{l-1,i}$  for $l = 1,2,3,\ldots$

$$s_{l,u} = w_{l,u,0} + \sum_{i=1}^{N_{l-1}} x_{l,i} w_{l,u,i} + d_{l,u}$$

$$y_{l,u} = \begin{cases} 1, & \text{if } s_{l,u} > 0 \\ 0, & \text{otherwise} \end{cases}$$

$\Delta y_{l,u} = y_{l,u}^{now} - y_{l,u}^{previous}$

$w_{l,u,i} \mathrel{+}= \rho \Delta r \Delta y_{l,u} (x_{l,i} - 0.5)$, for $i > 0$

$w_{l,u,0} \mathrel{+}= \rho \Delta r \Delta y_{l,u}$

The above equations for updating the hidden layer units are based on Williams' [11] definition of his non-episodic REINFORCE algorithm, given by:

$$\Delta w = \alpha(r - \bar{r})(y - \bar{y})$$
$$\bar{r}_t = \gamma r_{t-1} + (1-\gamma)\bar{r}_{t-1}$$
$$\bar{y}_t = \gamma y_{t-1} + (1-\gamma)\bar{y}_{t-1}$$

Williams describes these equations as pertaining to a bias weight having a constant input of 1. With the addition of variable inputs, $x_j$, and using $\gamma = 1$, these equations reduce to

$$\Delta w_j = \alpha(r - r_{t-1})(y - y_{t-1})(x_j - 1/2),$$

which correspond to the weight update equations given above for the hidden layer units. Letting $W$ represent all of the weights in the network, Williams proves for his REINFORCE class of algorithms, that the inner product of $E\{\Delta W|W\}$ and $\nabla E\{r|W\}$ is nonnegative. While this is not a convergence proof, it does prove that the change in weight values, $\Delta w$, tends to increase the expected value of $r$.

**Single flux quantum circuit implementation of the learning rules**

The weight update rules for the output layer can be implemented with standard rapid single flux quantum (RSFQ) digital logic gates [12,13] and a learning clock signal whose only requirement is to arrive any time after the inference output. While there are many potential ways to implement the logic, we used the following basic approach. Weight update values that are potentially zero, for example ($y^*$-$y$), are used as the clock signal for the logic gates that are closer to the synapse. The absence of this clock ensures that there is no unwanted weight update. The sign of any prefactor is determined as close as possible to the output layer. For example, $\Delta r \Delta y$ is determined at the soma output where $y^{now}$ is stored during inference. The final sign of the update is determined at the synapse since it depends on the input $x$ to the synapse. Finally, all values that are stored from the inference step are reset at the end of the learning cycle to prepare for the next inference cycle. Approaching the update logic in this way concentrates the number of logic gates at the output and soma level, minimizing the number of gates at the synapse and the amount of information that must be broadcast to each synapse.

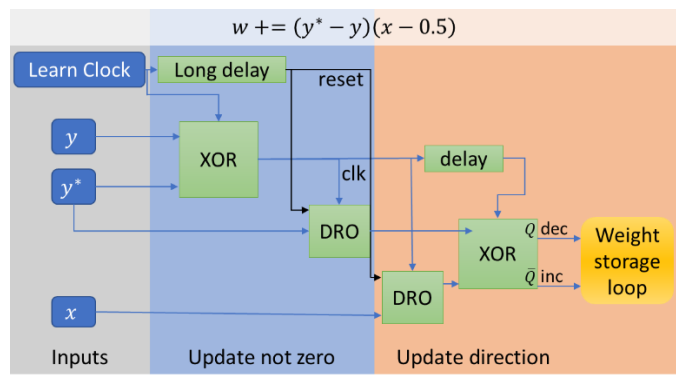

Figure 1: Block diagram of weight update logic for a synaptic weight in the output layer

Figure 1 shows an example of the nominal weight update logic for an output layer synapse. The desired output $y^*$ and network output $y$ are fed into an exclusive or (XOR) gate to determine if there is a weight update. If the values are the same, $(y^* - y) = 0$ there should be no update. Otherwise, the signal from this XOR is used as a clock for the rest of the update logic. To determine the sign of the weight update, the input to the synapse $x$ and the desired output $y^*$ are fed into an XOR gate via destructive readout (DRO) gates to determine the direction of the update. The output and complementary output of this XOR feed a single flux quantum (SFQ) pulse into the decrement or increment side respectively of the weight storage loop described below in the synapse section. A similar logical flow is followed for the other weight updates.

**Bipolar synapse and soma circuit**

While the reinforcement learning rules described above can be implemented with mostly standard digital RSFQ logic gates, the weight and threshold functions are implemented using an analog circuit that we break up into the synapse (weighting operation) and soma (summation operation). [5,14] The synaptic subcircuit in fig 2a is composed of a bipolar "synaptic" weight, meaning that the weight can be either positive or negative, connected to a summation "soma." The soma contains a line of mutually coupled inductors with a resistor to ground at one end and the input to a Josephson transmission line at the other. The inputs are set to arrive at the same time where they are weighted, and a portion of the incoming spikes are coupled into the soma. The resistor to ground at the bottom of the soma can be used to adjust the L/R time constant of this summation operation to account for wiring and jitter on the incoming signals, relaxing the timing condition on spikes arriving to the soma. The top of the soma feeds a biased JJ, which then applies the threshold activation function defined above as $y_{L,u}$. One technical difference is that in the circuit implementation, the threshold value is shifted from zero current in the soma to a positive current value. The threshold is set by a combination of the fixed bias current of the JJ at the top of the soma that is the input to the Josephson transmission line, and the $w_0$ bias weight, which is adjusted by the reinforcement learning logic. The effect however is the same; if the threshold is exceeded a single flux quantum (SFQ) pulse is emitted by the soma (1), otherwise nothing is emitted (0).

A more detailed look at the synapse shown in fig 2a follows: The weighting is accomplished using an inductive divider, where one path passes through an adjustable superconducting quantum interference device (SQUID) to ground, and the other path passes through an inductor in series with a resistor to ground. This second path's inductor is mutually coupled to the soma along with the other synapses that are connected to that soma. The resistor on the second leg of the divider is used to prevent unintended current buildup and to set the L/R time constant (leak rate) of the input signal to a synapse. The weight storage loop (shown in light blue wiring), controlled by the learning logic, affects the inductive divider by changing the amount of current circulating in adjustable SQUID, which in turn changes its inductance. Each synapse has a positive and negative branch. Both branches have the same inductive divider topology described above but with the coupling directions changed.

Figure 2b shows the behavior of the synapse from SPICE simulations. The storage loop starts at zero current and an input pulse $x$ is applied to the synapse. In this case the positive and negative branches have equal and opposite current coupled into the soma, which is seen as no spike appearing in the soma at the bottom of fig. 2b. Note that while perfect cancelation is accomplished in the simulation, it is not required to train the network. After the first input spike, a series of 32 SFQ pulses is applied to the decrement side of the weight storage loop in about 1 ns. Note that the number of SFQ pulses stored in the weight storage loop has a maximum of 30 for the simulated circuit shown here with a storage inductor of 250 pH. When the maximum number of SFQ pulses for the weight storage loop is exceeded, the additional SFQ pulses are expelled via the JJ on the other side of the loop and the maximum current value in the weight

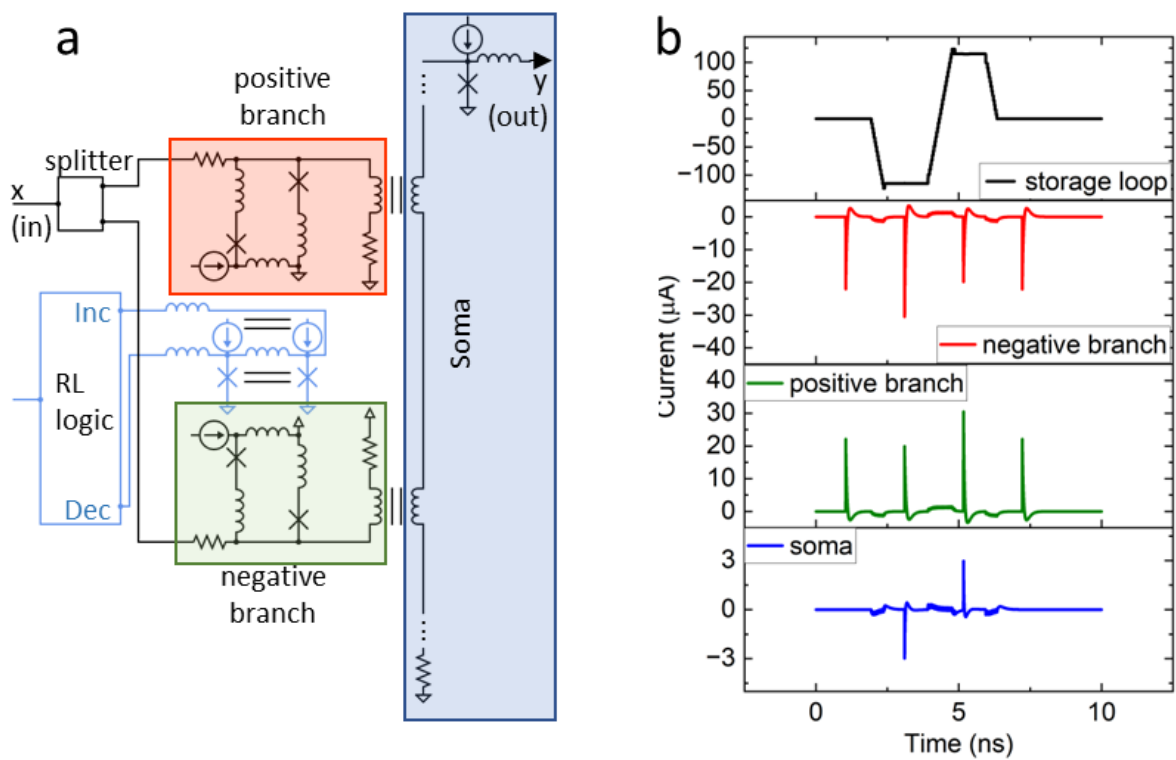

*Figure 2(a) bipolar synapse conceptual circuit diagram. Synapse portion in red box for positive weight, synapse portion in green box for negative weight. Blue wiring for the weight adjustment controlled by the reinforcement learning (RL) circuits described below. The two parts of the synapse are both coupled into the soma circuit, which can couple together multiple synapses. (b) SPICE simulation results showing the response of the synapse branches and soma to input spikes that occur when the storage weight is at zero, minimum, maximum, and after returning to zero.*

storage loop is maintained. After this minimum value of the weight storage loop is reached, a subsequent input pulse $x$ is applied to the synapse at 3 ns and now a larger portion of the negative branch and a smaller portion of the positive branch are coupled into the soma. This is seen in the soma at the bottom of fig. 2b as the net negative spike. After this, 64 SFQ pulses are applied to the increment side of the weight storage loop. Subsequently an input pulse $x$ is applied to the synapse, which results in a larger positive and smaller negative spike. This is seen in the soma as a net positive spike from the input *x*. Finally, 30 pulses are applied to the decrement side of the weight storage loop to return the weight storage loop to near zero current. The final input pulse *x* is applied to the synapse and once again the positive and negative branches cancel out and no spike is observed in the soma.

One other element that is used is a $w_0$, which acts as an adjustable bias for the soma. The $w_0$ circuit is the same as the positive branch of the bipolar synapse circuit. Each soma circuit has one such $w_0$ to adjust the threshold level. There is a separate weight loop and its own learning rule logic to set and store this $w_0$ for each soma. In addition, there is a $w_0$ input that spikes at every inference cycle. This results in the bias value being independent of whether an input signal was sent to a particular neuron, and no external adjustment of the soma bias is needed as described in the network equations above.

The weight storage is a superconducting loop with a biased JJ on either side, shown in light blue in the middle of fig. 2a. Current can be injected in either direction in this loop following the result of the learning rules. The size of the inductor sets the effective bit depth of the weight storage. In addition, the adjustable SQUIDs in the synapses are set so that they always remain in the non-spiking state regardless of the weight value, which results in a smooth variation in weight value. In the simulations described in this article we use a 250 pH storage inductor, which results in just under a 5-bit weight. The bit depth is approximately linear with inductor size, meaning that one can use a 500 pH storage inductor for a weight depth of about 6 bits.

## Results

### Examples of small-scale functioning network

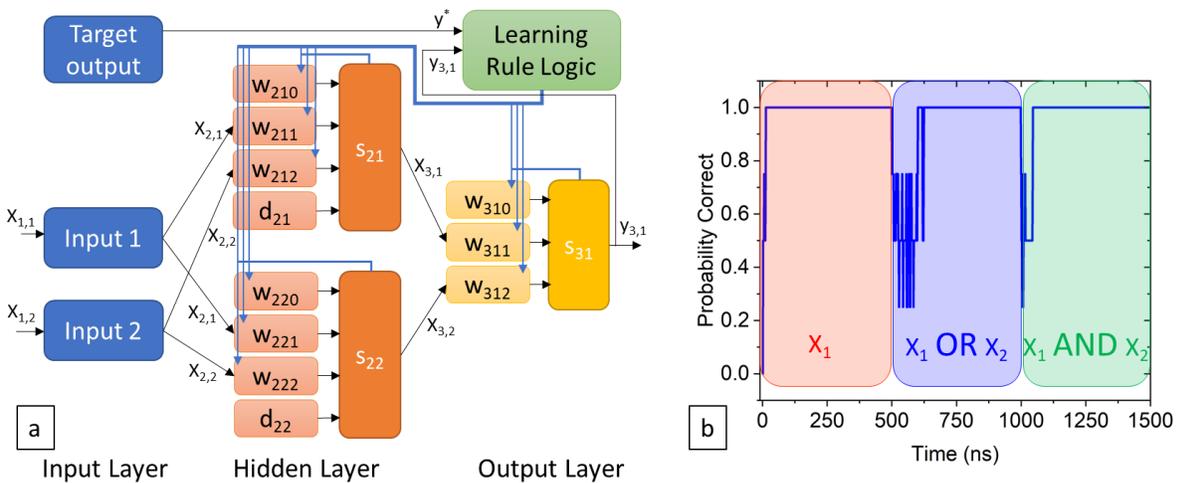

*Figure 3a Block diagram of small scale demonstration network. (b) Probability correct versus time for a continuous SPICE simulation of the network learning three different functions. The functions are spike when input 1 is 1 for the first 500 ns, spike when either input 1 or input 2 spikes for the next 500 ns, and spike only when input 1 and input 2 spike for the last 500 ns.*

We constructed a small SPICE network using the bipolar synapse, soma circuits, and the digital SFQ learning logic circuits described above to implement the reinforcement-based learning rules.

Figure 3a shows a basic block diagram of this network. There are two inputs that each take an external binary signal and convert it to an SFQ spike for a 1 and no spike for a 0. In addition,

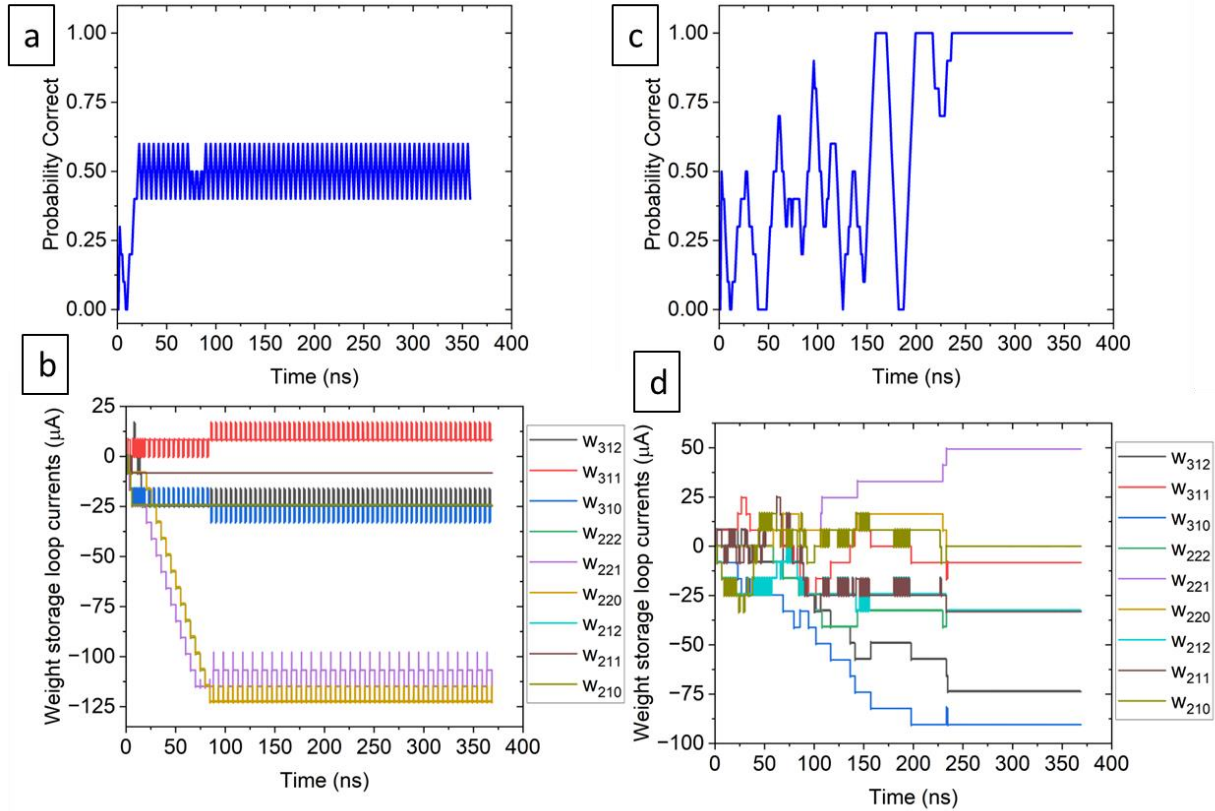

*Figure 4(a) probability correct vs. time for the network attemting to learn the X2 function with a problematic input sequence and 1% stochastic excitation magnetude. (b) the weight storage currents (offset for clarity) of the synapses for the simulation in (a). (c) probability correct vs. time for the same network and input sequence as in (a) and (b), but with a stochastic exication magnetude of 20 % of the maximum synapse value. (d) the weight storage currents (offset for clarity) of the synapses for the simulation in (c).*

there is an input clock (not shown in the diagram) that spikes on every input cycle and is used for both $w_0$ and the stochastic excitations.

The input layer is fully connected to a hidden layer of two units, which is subsequently fully connected to an output layer with one unit. Each node in the block diagram of fig. 3a contains all the synapses required to weight the inputs as well as the soma circuit that performs the summation and activation. In addition, the $w_0$ bias and stochastic excitation $d$ are included in the nodes and are summed within the same soma. Figure 3b shows the results from a single SPICE simulation of the network. The network is trained to classify an input vector by providing the desired output $y^*$ for a series of input vectors. In this simulation input 1 had an input series of 0011 repeated and input 2 had an input series of 0101 repeated. The time for inference plus learning in this example was 1 ns. The target output was divided into three phases. In phase 1, which covers the first 500 inputs (500 ns in time), the network learned to spike whenever input 1 spikes ($X_1$) regardless of input 2. In the second phase (from input 501 ns to 1000 ns) the network learned to spike when either input 1 or input 2 spiked (OR). In the third phase, from 1001 ns to 1500 ns, the network learned to spike only when input 1 and input 2 spiked (AND). The effective training of these three phases can be seen in the probability correct of fig. 3b. The probability correct is a running average of four cycles of the probability that the output was correct for a given input vector, meaning that a value of 1 implies that the output of the network matched the target output for at least four cycles. Figure 3b shows that given a new desired output, the network can adjust its own weights on the fly to learn this function.

The previous functions could be robustly learned without stochasticity. However, most non-trivial classifications benefit from the enhanced weight exploration that the additional stochastic term in the hidden layer nodes provides. Figure 4 shows the same network training to spike whenever input 2 spikes ($X_2$). In this example, input 2 was fed an alternating signal of 010101… Without providing the same input for two samples in a row $\Delta r$ and $\Delta y$ no longer provide a well-defined reinforcement signal for the weight updates to follow. Figure 4a and 4b show the probability that the network outputs the correct value and the stored weights when training with the pathological input pattern. The periodic oscillation of the value of input 2 leads to an oscillation of the weight values around a local minimum in the error, which prevents the network from converging on the correct output behavior.

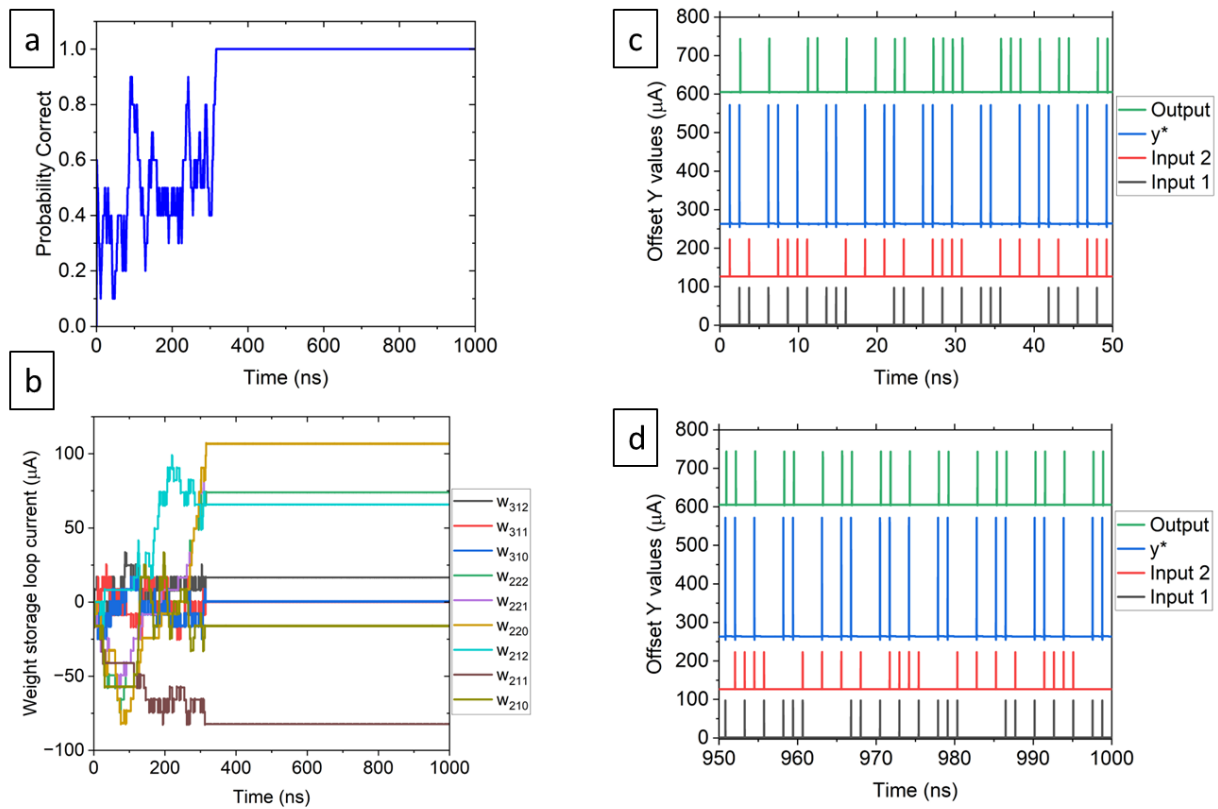

*Figure 5(a) the probability correct versus time for a small network learning the XOR function. (b) the weight storage currents (offset for clarity) of the synapses in this network vs. time. (c) spiking patterns for the input and output of the network at the start of training. (d) spiking patterns for the inptu and output of the network after the network has learning the XOR funciton.*

A stochastic excitation $d_{L,u}$, defined in the learning rules above, can enable training around this flawed input sample. Figures 4a and 4b have a stochastic excitation of the hidden layer units that was set to <1% of the maximum value of a synaptic weight.  Figures 4c and 4d have the same input pattern and network as figs. 4a and 4b, but now have the stochastic excitation of the hidden layer units set to have a magnitude of approximately 20 % of the maximum weighted input value.  As can be seen in fig. 4c the network now correctly learns to identify input 2 regardless of the input 1 value.  In general, we find that a stochastic excitation of between 20 % and 50 % of the maximum weighted input is an effective magnitude for weight space exploration. It should be noted that if the inputs were sampled twice, *e.g.* 00110011…, X2 could be correctly learned without the additional stochastic excitation, as expected.  Also, we note for completeness that all cases in Fig. 3 can be learned with a stochastic excitation of 20 % to 50 % of the maximum weighted input.  In general, as the problems become more complex the additional weight exploration provided by the stochastic excitation to the hidden units may become more beneficial for faster learning convergence. [15]

Figure 5 shows the same network as in figures 3 and 4 with a stochastic excitation of about 20% of the maximum weighted input learning the XOR function.  This is a nonlinearly separable problem, which therefore requires the hidden layer and demonstrates the generality of both the network and its training rules.  In this example we use an input pattern that is a set of 16 quasi-random input vectors that is repeated for 1000 total cycles.  Figure 5a shows the probability that the output correctly spikes when only input 1 or input 2 spikes, averaged over 4 cycles.  Figure 5b shows the weights of the synapses as the network correctly learns the XOR function in less than 400 ns.  Figure 5c shows the spiking pattern of the inputs, the desired output *y\**, and the actual network output for the first 50 ns of training.  Figure 5d shows the spiking patterns from 950 ns to 1000 ns of the SPICE simulation where one can see that the network has correctly learned the XOR function.

**Extension to MNIST:**

To investigate the suitability of the reinforcement learning approach for larger problems, simulations were run with a python model implementing the learning rules described above applied to the problem of classifying images of handwritten digits in the modified national institute of standards and technology (MNIST) dataset. [16] We tried to put limits in the simulations to reflect the constraints of the SPICE models above, these include a weight bit depth of 5 bits and a fan-in limitation of 50.  However, details such as cross talk and wiring layout were not considered.  The MNIST (Modified National Institute of Standards and Technology) dataset consists of 70,000 images of handwritten digits that is often used as a benchmark for comparing image classification algorithms. [17] Each image is a 28 x 28 image of 784 pixel intensities. Given these intensities as inputs, a neural network was trained to classify the input image as the correct digit. The output consisted of 10 values between 0 and 1 indicating the probability that the input image is digits 0 through 9.

The results in Table 1 were obtained with neural networks containing zero, one, two, or three hidden layers, each with 200 units. 60,000 images were used for training the neural networks and the remaining 10,000 images were used to test the performance of the trained networks. The networks were trained for 500 epochs. Accuracy is determined by the percentage of image samples correctly classified.

| Number of Hidden Layers | Training Set Accuracy | Test Set Accuracy |
|---|---|---|
| 0 | 14.8 % | 14.6 % |
| 1 | 87.2 % | 87.0 % |
| 2 | 90.6 % | 90.1 % |
| 3 | 90.2 % | 90.6 % |

Table 1. The percentage of train and test set images that are correctly classified by networks having no, 1, 2, or 3 hidden layers of 200 units each.

The current hardware design of the soma has a limit of 50 incoming connections. [18] This was modeled in the python implementation in the following way. Each unit in the first hidden layer was connected to a randomly chosen 7 x 7 patch of the 28 x 28 original image. While this is like the kernels in convolutional neural networks, here convolution is not performed. Each unit only gets the 49 intensities from the single 7 x 7 patch randomly assigned to it when the neural network is defined. The units in the remaining hidden layers receive inputs from 49 randomly selected outputs from the previous hidden layer. Each of the 10 units in the output layer receives 49 randomly selected outputs from the last hidden layer. Thus, each unit in the neural network has 49 weights corresponding to its 49 inputs, plus one bias weight, for a total of 50 weights in each unit.

Classification is performed by picking the unit with the largest value among the 10 units in the output layer. Several possibilities exist for implementing this "argmax" operation in the circuit, including a winner-take-all operation based on recurrent connections in the output layer. SPICE modeling of this part of the circuit is beyond the scope of this paper and will be investigated in future work.

## Discussion

### Scaling

In the above examples we show SPICE simulations of a small proof-of-principle self-training network that fully implements reinforcement learning using a common global reward signal. The network is a mixed digital and analog set of circuits that is fully self-contained. The bipolar synapse and soma constitute the analog portion of the network. The analog weights are adjusted via injection of SFQ pulses into a superconducting weight storage loop. SFQ pulses can be injected on either side of the loop, resulting in the ability to increase or decrease the strength of the weight in either direction. The bipolar synapse architecture further enables any synapse to be either inhibitory or excitatory in nature. This bipolar synaptic network style enables a more efficient hardware implementation since synapses do not need to be duplicated or assigned a sign in advance of the training. Because the learning rules determine the direction of a given weight update, there is no need to precisely set the weight value or threshold in the analog portions of the network. This alleviates one of the main challenges in the use of analog hardware, since the typical process variations inherent in all hardware can be accommodated by the learning rules.

In the network described above with 2 input units, 2 hidden units and 1 output unit, the time to run a full inference and training cycle is about 1 ns. This network architecture is scalable to much larger networks. In general, the per layer inference time will be set by the choice of resistors in the synapse and soma circuits, both of which set an L/R time constant. In addition, the fanout required for the input as the layers get wider will also add to the per-layer inference time. In our simulations we chose a time constant of about 50 ps to account for typical jitter since the network assumes that all inputs from the previous layer arrive at the soma at the same time within about 1 time constant. Different time constants can be chosen to run faster or relax the timing constraint depending on the desired goal. In general, a 50 ps time constant would imply a per-layer timing delay of about 150 ps (i.e., assuming a three time constants decay). In our 2 unit wide network the additional delay from fanout is roughly an additional 6 ps. Fortunately, the time scaling for fanout is logarithmic so even at a width of 10,000 units the time delay added by fanout to all units is about 84 ps. Finally we note that as networks get larger, the requirement that all signals arrive together for a layer can be relaxed to all signals arriving together for a given soma. The output of the layer can then be resynchronized before moving to the next layer with minimal additional circuitry and timing delay.

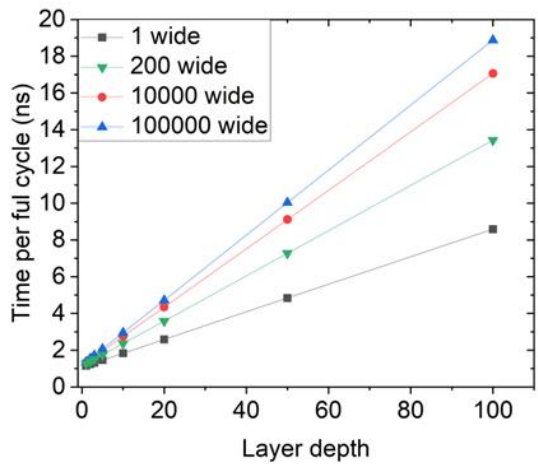

Figure 6  Total time delay for inference and learning plus weight update vs. the number of layers in a network for different layer depths.

Learning time scales favorably in this architecture in part because the weight updates can be globally asynchronous. In our non-optimized circuit, it takes about 250 ps to determine the reward signal. It takes an additional 250 ps to determine the soma update for each unit and another 250 ps for the synapse update at each synapse. Thus, it takes about 750 ps from the final output until the synapses are all updated. Since the update rules do not need to be applied sequentially, the reward signal can be broadcast back to all units at the same time. Thus even for a 1,000,000 unit network the additional fanout time overhead is about 120 ps. This gives us a total learning time of about 1 ns with minimal impact on the depth of the network.

Figure 6 shows the total time delay for a full inference plus learning cycle versus the number of layers in the network for four different network widths. There is a linear scaling of the inference latency on the number of layers, which dominates the total time as the network grows in size. However, the total time for a learning cycle is still relatively short because of the ability to broadcast the reward signal to the entire network at the same time. While the wiring for this type of network will be a challenge, it is worth noting that in superconducting SFQ circuits there are no voltages to raise on the wiring layers and no associated RC time constants, as there would be in more conventional CMOS based technologies. Superconducting neuromorphic computing can take advantage of this difference and implement wiring that behaves closer to that of photonic waveguides than conventional CMOS wiring. It is worth noting that there will be additional timing delays as the network grows due to wiring, but these have a minimal impact on the time scaling. The time to propagate SFQ pulses in Nb is roughly c/3 (where c is the speed of light in vacuum), or 10 ps/mm. [19] While repeater JJs will be needed for JTLs, passive superconducting transmission lines can be used for longer wires. Thus, the additional timing overhead of routing the broadcast signal should be much less than 1 ns. This gives us a 1 million node, 100 layer deep learning time of about 31 ns or about 30 million cycles per second and favorable time scaling as the network gets larger.

Finally, it is worth noting that million node networks are not within reach of current fabrication densities. However, our unoptimized example network contained a little less than 1000 JJs for 5 nodes in total. Using 200 JJs per node as a very rough estimation, networks with order 1000 nodes should be achievable with current fabrication technology. This includes the 3 hidden layer MNIST network in our python simulations. If implemented in our JJ reinforcement learning architecture, the 3 hidden layer MNIST network would take about 100 ms to train, after which image classification could be run at a rate of 1 image every 150 ps (6.7 GHz).

**Summary/conclusion**

In this article we have shown a proof of principle implementation of reinforcement learning in a mixed analog and digital neuromorphic SFQ architecture. The learning rules were constrained to be relatively efficient to implement with a spiking JJ neural network. We use SPICE simulations to show the proof of principle for a small scale non-linearly separable problem (XOR), as well as the effectiveness of stochastic weight exploration in the hidden layer units. We have extended the same reinforcement learning logic demonstrated in the SPICE simulations to python where the network can be scaled to perform MNIST level classifications. The highlights of this architecture include the ability for the network to self-train while only requiring the addition target output value $y^*$. In addition, the fast-learning time leads to a cycle time of about 1 ns for small networks and preferable time scaling even to large networks. In addition, after the learning phase, the network can be run in inference only mode. When operating in inference only mode the latency of the initial result will scale as shown above. However, one does not need to wait for the final output before the next input. This can lead to large scale networks that can perform complete classifications at rates faster than 5 GHz.

 Acknowledgements

We thank Dr. Bryce Primavera and Dr. Andrew Dienstfrey for their helpful discussions.